# Study of the influence of the molecular organization on single-layer OLEDs' performances

Laurent Aubouy, Philippe Gerbier*, Christian Guérin

Laboratoire de Chimie Moléculaire et Organisation du Solide (CNRS, UMR 5637), Université Montpellier II, CC007, Place Eugène Bataillon, 34095 Montpellier Cedex, France.

Nolwenn Huby, Lionel Hirsch, Laurence Vignau

Fédération de Recherche SyMe (CNRS, FR 2648), Université Bordeaux 1, 351, Cours de la Libération, 33405 Talence Cedex, France.

Corresponding author. Tel. : +33 (0)4 67 14 3972; fax: +33 (0)4 67 14 3852.

E-mail address: gerbier@univ-montp2.fr

**Abstract**

We have synthesized three new silole derivatives with incrementally flexible structure to tune their packing ability and therefore study the influence of the molecular organization in single-layer OLEDs. This architecture was chosen since the absence of organic layers interfaces allows a better evaluation of the role of the molecular arrangement in the active layer. The examination of the EL properties gives evidences of the prominent role of the molecular organization on the OLED efficiency. A crystalline-like organization of the molecules allows high current density but low luminance efficiency since an excessive electron current flow is involved compared to the hole one, and the recombination rate is poor. On the contrary, disordered assemblies of molecules allows better performances by avoiding unfavourable π-stacking, while keeping good intermolecular orbital overlaps to support charge carrier transport.

## 1. Introduction

Since Tang and VanSlyke made shine the first Organic Light Emitting Diode (OLED) [1], much efforts have been made to improve these devices. Thus, multi-layers [2], doped [3], and phosphorescent [4] OLEDs have been widely described and showed recently their limit in terms of durability because of interfacial or phase separation problems [5-7]. Along with the implementation of further architectural refinements, the enhancement of the electroluminescence (EL) efficiency passes obviously through the synthesis of the new molecules possessing very dedicated capacities [8]. For instance, good quantum yield, high charge mobility and adapted HOMO-LUMO levels are generally required to achieve good device output [9,10]. Moreover, recent studies have also demonstrated the role of molecular organization on the luminescent properties [11]. With this in mind, we have designed a series of emissive silole derivatives having both the required electron- and hole-transporting properties. These molecules were chosen since they intrinsically exhibit a good chemical stability along with optoelectronic and electron–transporting properties originating from an effective interaction between the σ*-orbital of the silicon atom and the π*-orbital of the butadiene moiety [12]. The missing hole-transporting properties were added by just grafting arylamine side-groups to the central silole ring. As a result, single layer OLEDs based on these molecules show high performances, well above those observed when $Alq_3$ is used as the active layer [13, 14]. We will report herein the implementation of a series of siloles having incrementally varied structures, π-conjugation, and intramolecular overlap properties. The main objective being the demonstration of some of the relationships that may exist between the molecular arrangements and the OLEDs efficiency.

**2. Experimental**

*2.1. Materials and methods*

Solvents were distilled prior to use. THF and ether were dried over sodium/benzophenone, and distilled under Argon. All the reactions were carried out under argon atmosphere. $^{1}$H, $^{13}$C and $^{29}$Si NMR spectra were recorded on a Bruker Advance 200 DPX spectrometer, the FT-IR spectra on a Thermo Nicolet Avatar 320 spectrometer, the UV-visible spectra on a Secomam Anthelie instrument and the MS spectra on a Jeol JMS-DX 300 spectrometer. Synthesis and characterization of compounds **1, 5a, 6a, 7a** were previously described [11]. Organic glasses preparation for DSC studies: the sample of silole was heated above its melting point and cooled at a rate of 30°C/min to room temperature.

Devices of ca. 10 mm$^2$ were fabricated on ITO-coated glass substrates (Merck, thickness ≈ 115 nm, sheet resistance ρ ≈ 17 Ω/□). A 50 nm-thick layer of PEDOT-PSS was spun at 5000 rpm on top of ITO and baked at 80°C for about 1 hour. On the PEDOT-PSS layer, all the organic compounds so as the cathodes were thermally evaporated under secondary vacuum (10$^{-6}$ mbar) at about 1 nm/s to a thickness of 60nm. Finally, a 80 nm-thick Calcium capped by an Aluminum layer was evaporated through a shadow mask on top of the silole derivative. Current-voltage characteristics were recorded using a Keithley 2400 Sourcemeter and luminance-voltage with a photodiode placed under the OLED and coupled to a HP multimeter. Electroluminescence (EL) spectra were measured using an Ocean Optics HR2000 CCD spectrometer. All electroluminescent devices were kept and characterized in a glove box under nitrogen.

*2.2. Syntheses*

2.2.1 p-2,2'-dipyridylaminophenyl-4-bromophenylether (**2**).

A mixture of bis(4-Bromophenyl)ether (7.14 g, 21.7 mmol), di-2-pyridylamine (1.5 g, 8.7 mmol), $K_2CO_3$ (1.4 g, 10.4 mmol) and $CuSO_4$ (0.217g, 0.87 mmol) in water (20 mL) and $CH_2Cl_2$ (100 mL) was stirred well and evaporated to dryness in vacuum. The mixture was ground in a mortar and heated in a sealed tube at 205 °C for 6 h. After being cooled at room temperature, the mixture was dissolved in $CH_2Cl_2$ (100 mL) and water (100 mL) and extracted three times with $CH_2Cl_2$. After evaporation of the solvent, the residue was subjected to column chromatography $CH_2Cl_2$/THF (95/5) to afford compound 2 (yield: 83%). Mp : 102°C. $^1$H NMR (CDCl$_3$, δ, pmm) : 8.35 (dd, $J_1$ = 7, $J_2$ = 2Hz, 4H), 7.60 (td, $J_1$ = 7, $J_2$ = 2Hz, 4H), 7.47 (d, J = 9Hz, 2H), 7.21 (d, J = 9 Hz, 2H), 7.06-6.81 (m, 4H). $^{13}$C NMR (CDCl$_3$, δ, ppm) : 157.99, 156.17, 154.50, 148.41, 140.33, 137.63, 132.74, 128.92, 120.82, 119.84, 118.15, 116.71, 115.95. HRMS (fab+, m-nitrobenzyl alcohol matrix) m/z: calcd for (M+) $C_{22}H_{16}N_3BrO$ 418.0555, found 418.0547.

*2.2.2. General procedure for the preparation of siloles **8, 9, 10**.*

To a solution of compounds **1** or **2** in THF (30 mL) at -78°C, 1.3 equivalent of nBuLi 2,5M in hexane were added. After 1 hour under stirring, a solution of 1.3 equivalent of B(OMe)$_3$ in THF (20 mL) was added and the reaction left overnight at room temperature. Then the mixture was hydrolysed with a saturated solution of NH$_4$Cl, extracted with ethyl acetate (50 mL). The organic layer was washed with water, dried with anhydrous MgSO$_4$ and evaporated under vacuum. The white residue is then subjected to a column chromatography on silica gel with $CH_2Cl_2$/MeOH (95/5) as eluant to afford the desired boronic acid **3** or **4** as white solids. (50%). In a round bottom flask, dibromosiloles **7a** or **7b** were dissolved in toluene (30 mL), and Pd(PPh$_3$)$_4$ (0,1 eq) is the added. After stirring during 10 min, desired Boronic acid (4 eq) in ethanol solution (10 mL), and NaOH (10 eq) in water (10 mL) were added. The reaction left under reflux during 72h, and was then cooled to room temperature, then is diluted with 100 mL of water and 100 mL of CH$_2$Cl$_2$. The organic layer is separated, dried and evaporated under vacuum.

*2.2.3. 1,1-Dimethyl-2,5-bis(8-para-biphenyldi-2-pyridylamine) silole (**8**).*

The obtained solid is purified by column chromatography (Silicagel) using CH$_2$Cl$_2$/THF (90/10) as eluant to afford **8** as a bright yellow solid (37%). Mp: 270°C. $^1$H NMR (CDCl$_3$): 8,45, dd, J$_1$ = 6, J$_2$ = 2 ; 4H ; 7,62-7,54, m, 16H ; 7,42, s, 2H ; 7,29, m, 4H ; 7,09-6,99, m, 8H ; 0,63, s, 6H. $^{13}$C NMR (CDCl$_3$): 157,49 ; 148,15 ; 144,34 ; 143,41 ; 138,85 ; 138,65 ; 138,60 ; 137,93 ; 137,88 ; 128,81 ; 128,62 ; 127,43 ; 127,33 ; 118,59 ; 117,01 ; -2,69. $^{29}$Si NMR (CDCl$_3$): 2,595. HRMS (m/z): calcd for [C$_{50}$H$_{40}$N$_6$Si]H$^+$ : 753,3162 found 753,3128.

*2.2.4. 1,1-Dihexyl-2,5-bis(8-para-biphenyldi-2-pyridylamine) silole (**9**).*

The mixture is successively submitted to a silica gel and a alumina columns using CH$_2$Cl$_2$/THF (88/12) and CH$_2$Cl$_2$/MeOH (99,5/0,5) as eluants, respectively to afford **9** as a yellow solid. (53%) Mp: 83-84°C. $^1$H NMR (CDCl$_3$): : 8,33, dd, J$_1$ = 6, J$_2$ = 2 ; 4H ; 7,70-7,60, m, 16H ; 7,49, s, 2H ; 7,30-7,25, m, 4H ; 7,14-7,02, m, 8H ; 1,35-0,81, m, 26 H. $^{13}$C NMR (CDCl$_3$): 158,49 ; 148,77 ; 144,88 ; 143,62 ; 139,69 ; 139,05 ; 138,84 ; 137,93 ; 137,82 ; 128,09 ; 127,83 ; 127,48 ; 127,14 ; 118,65 ; 117,44 ; 33,18 ; 31,72 ; 24,00 ; 22,88 ; 14,69 ; 13,64. $^{29}$Si NMR (CDCl$_3$): 3,645. Masse (FAB+, m-NBA) : MH$^+$, m/z = 893.

Anal. Calcd for C$_{60}$H$_{60}$N$_6$Si : 80,67 %C, 6,77 %H, 9,40 %N Found: 78,48 %C, 6,79 %H, 9,12 %N.

*2.2.5. 1,1-Dimethyl-2,5-bis(8-para-diphenylether-di-2-pyridylamine)silole (**10**).*

The mixture is submitted to a column chromatography (Silicagel) CH$_2$Cl$_2$/THF (85/15) to provide **10** as a bright yellow solid. (80%). Mp: 173°C. $^1$H NMR (CDCl$_3$): 8,39, dd, J$_1$ = 6, J$_2$ = 2 ; 4H ; 7,66-7,58, m, 18H ; 7,41, s, 2H ; 7,25-6,94, m, 20H ; 0,62, s, 6H. $^{13}$C NMR (CDCl$_3$): 158,29 ; 156,81 ; 155,57 ; 148,71 ; 144,71 ; 140,155 ; 139,31 ; 138,40 ; 138,15 ; 136,55 ; 131,29 ; 129,40 ; 128,59 ; 127,61 ; 127,01 ; 120,28 ; 119,89 ; 118,49 ; 117,57 ; -3,52. $^{29}$Si NMR (CDCl$_3$): 2,539. Masse (FAB+, m-NBA) : MH$^+$, m/z = 936. Anal. Calcd for C$_{62}$H$_{48}$N$_6$SiO$_2$ : 79,45 %C, 5,16 %H, 8,90 %N Found: 77,23 %C, 5,64 %H, 7,96 %N.

*2.3. Computational procedures.*

The ab initio computations of molecules **8** and **10** were carried out with the Gaussian 03 package of programs at density-functional theory (DFT) level using Pople's 6–31G split valence basis sets supplemented by d-polarization functions [19]. DFT calculations were carried out using Becke's threeparameter hybrid exchange functional [23] with Lee–Yang–Parr gradient corrected correlation functional (B3LYP) [24, 25]. Thus, the geometries were optimized with B3LYP/6–31G and the energies and electronic structures were then calculated for single point at the B3LYP/6–31G* level. Restricted Hartree–Fock formalism was applied for geometry optimization and electronic structures calculations, no constraints of bonds/angles/dihedral angles were applied in the calculations and all the atoms were free to optimize.

### 3. Results and discussion

*3.1. Synthesis*

Originally synthesized by Braye and Hubel in the 60's [15], siloles experienced a wonderful rebirth in the 1990s with the implementation of new efficient synthetic methods allowing the introduction of various functional groups at 2,5-positions of the silacycle [16, 15]. However this procedure requires the presence of additional phenyl ring at the 3,4-positions, whose presence is detrimental to the luminescence quantum yield. This reason led us to consider the 2,5-diarylsiloles series though they were less described due to their difficulty of synthesis. We reported in a previous communication a new synthetic route that allows the convenient preparation of siloles **6a** and **6b** that serves as starting material for a series of new 2,5-diarylsiloles (Scheme 1) [14]. The linkage of the hole-transporting group was carried out through palladium-catalyzed Suzuki's cross-coupling reactions. Thus, bromine-substituted hole-transporting synthons **1** and **2** were synthesized by an Ullman's condensation in sealed tubes with an average yield of 83%. They were converted to the corresponding boronic acids **3** and **4** by lithiation with n-BuLi followed by the addition of $B(OMe)_3$. The hydrolysis, gave the boronic acids were used in the Suzuki's cross-coupling reaction in presence of $Pd(PPh_3)_4$ and sodium hydroxide. Siloles **8, 9** and **10** were obtained with yields of 37, 80 and 53%, respectively as bright yellow solids.

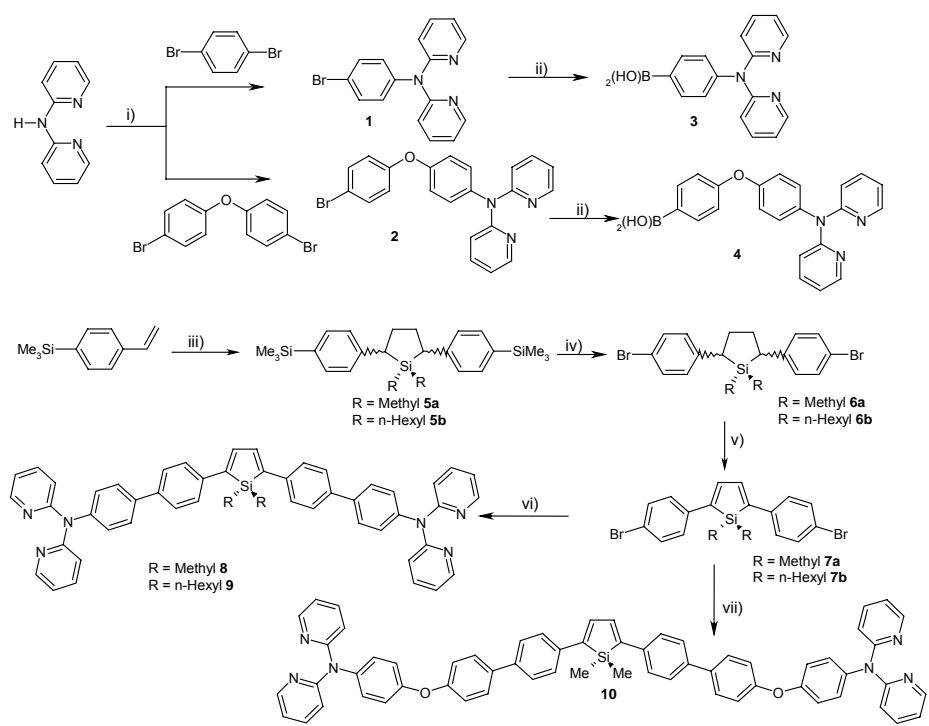

Scheme 1. (i) CuSO$_4$, K$_2$CO$_3$, 205°C, 6h ; (ii) 1) *n*-Buli and 2) B(OMe)$_3$ ; (iii) Li, R$_2$SiCl$_2$ (R=Me, *n*-Hex); (iv) Br$_2$ ; (v) 1) Benzoyl peroxide, N-bromosuccinimide and 2) CH$_3$CO$_2$H/CH$_3$CO$_2$K ; (vi) **3**, Pd(PPh$_3$)$_4$, NaOH ; (vii) **4**, Pd(PPh$_3$)$_4$, NaOH.

*3.2. UV-visible absorption and photoluminescence properties*

UV-visible absorption and fluorescence spectra have been measured in diluted dichloromethane solutions. All the spectra are characterized by the presence of two intense absorption bands in the 280 and 420 nm regions (Table 1), the emission spectra show an emission peak around 515 nm. As discussed previously, the absorption band observed at *ca.* 280 nm originates from π-π* transitions of the aryl groups, whereas the absorption band observed at *ca.* 420 nm is characteristic of π-π* transitions in the silole ring [14]. When compared with the values reported for 1,1-dimethyl-2,3,4,5-tetraphenyl silole **TPS** (absorption: $\lambda_{max}$ = 359 nm, emission: $\lambda_{max}$ = 467 nm, $\Phi_{em}$ = 0.0014), and 1,1-dimethyl-2,5-

diphenyl silole **DPS** ($\lambda_{max}$ = 376 nm, emission: $\lambda_{max}$ = 463 nm, $\Phi_{em}$ = 0.29), the red-shift that is observed for all the new siloles both in absorption and emission is indicative of a substantial perturbation of the HOMO-LUMO levels. This perturbation is provided by both the more extended π-conjugation afforded by the biphenyl side-groups and by the presence of a terminal electronically active heteroatom (N, O). Moreover, as expected (see above), the removal of the phenyl rings at the 3,4-positions, associated to the extension of the conjugated backbone, involves a dramatic increase of the quantum yield that reaches values of 97% for **8** and **9** and 65% for **10**. The relatively low value of $\Phi_{em}$ that is observed with **10** originates from the disruption of the conjugation afforded by the bridging oxygen atom that lead, both the silole and the dipyridylamino chromophores to be considered as being electronically independent (see below). In this situation the intramolecular energy transfer from the dipyridylamine sides groups to the silole ring, which is also responsible for the observation of such high quantum yields, appears to be less efficient [18]. Finally, it is worthy to note that the absorption and emission properties of molecule **8** are identical as those of molecule **9**, indicating that the organic groups (methyl *vs* hexyl) presents on the silicon atom don't affect the electronic properties of the siloles.

3.3. Calculations

The gas-phase geometry optimization of siloles **8** and **10** was carried out using density functional theory (DFT) calculations with the B3LYP functional. Due to the size of the molecule, the geometry optimization was performed with the 6-31G basis set to the standard convergence criteria as implemented in Gaussian G03w software [19]. The electronic structures were calculated at the B3LYP/6-31G* level of theory. The geometry of siloles **8** and **10** virtually approaches a $C_2$ symmetry, even though the optimizations were done without constrains (Fig. 1). Both the siloles display a nearly coplanar arrangement of the adjacent phenyl rings (A) in respect to the central silole ring (S) with average torsion angles of 13.2° for **8** and 12.3° for **10**. The phenyl-phenyl average torsion angles between rings A and B are also very similar in both the cases with values of 33.0° for **8** and 35.4° for **10**. These values as well as the other structural parameters (bond lengths, bond angles and torsion angles) are well

Table 1. UV-visible absorption and photoluminescence (PL) data for siloles **8**, **9** and **10**.[a,b]

| Silole | $\lambda_{max}$ Abs | $\lambda_{max}$ PL | $\Phi_{em}$ |
|---|---|---|---|
| **8** | 419 (5.63) | 524 | 0.97 |
| **9** | 421 (5.62) | 524 | 0.97 |
| **10** | 416 (5.45) | 508 | 0.65 |

[a] $10^{-3}$M solutions in $CH_2Cl_2$. [b] $\lambda_{max}$ in nm, (log ε).

in the range of those found in the Cambridge Structural Database for related structures [20].

The HOMOs and LUMOs shown in Fig. 2 display qualitatively similar surfaces if we do not consider the dipyridylamine side-groups. The filled π orbitals (HOMOs) and the unfilled orbitals (LUMOs) are mainly dominated by orbitals originating from the silole ring and two biphenyl groups at the 2,5-positions, while the LUMOs have significant orbital density at two exocyclic σ bonds on the ring silicon, giving rise to the genuine σ*-π* conjugation of siloles [12, 17, 21]. Focussing now on the dipyridylamino side-groups, they are weakly involved in the LUMO levels for both the siloles. The most stricking difference is found in the HOMO levels since very small orbital density is found on the dipyridylamino side-groups of **10**, when compared with the HOMO orbital of **8**. Thus, the major consequence originating from the insertion of an atom of oxygen in the backbone, is the disruption of the π-conjugation

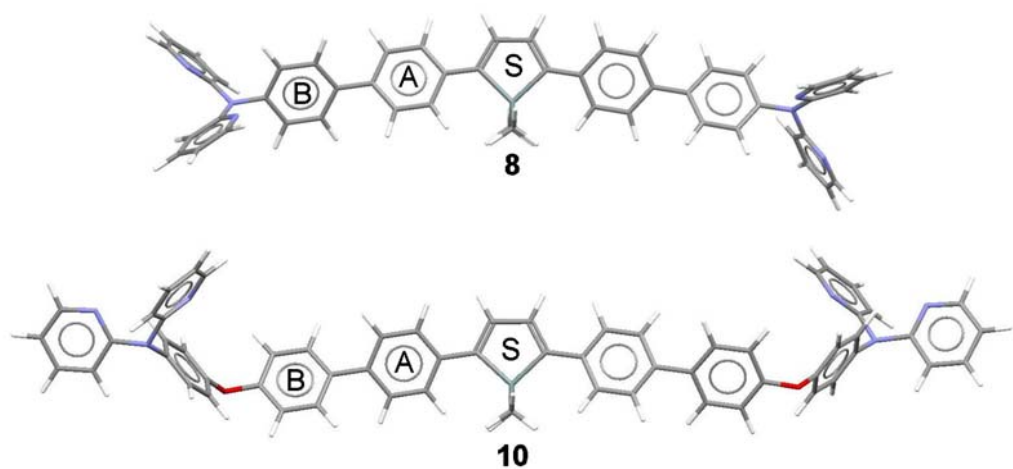

Fig. 1. B3LYP/6-31G optimized geometries of siloles **8** and **10**.

between the electron-transporting silole core and the hole-transporting dypirydylamino side-groups. One of the measurable effect of this disruption of conjugation is observed in both the HOMO and LUMO levels of **10** that are shifted to slightly higer energy when compared with

**8** (Fig. 2). As a consequence, the HOMO-LUMO gap is increase by ca. 50 meV. These results are well in line with those obtained from UV-Vis spectroscopic measurements.

*3.4. Molecular organization and thin film properties*

The molecular organization of vapor-deposited silole thin films (50 nm) onto PEDOT-PSS/ITO substrates were carried out by means of optical microscopy under polarized light, X-ray diffraction (XRD) and differential scanning calorimetry (DSC). Optical microscopy images (Fig. 3) show two different film morphologies: silole **8** forms highly featured film, whereas siloles **9** and **10** form rather homogeneous ones. The microscopic observation of the film formed with **8**, under polarized light, shows that the featured zones are highly organized, as attested by the observation of birefringence. Despite of our efforts no diffraction peak, related to the observed molecular organization, was observed in the XRD patterns. Concerning siloles **9** and **10** no birefringence was observed, indicating that the films are amorphous. All these observations are confirmed by the DSC studies (Fig. 4) carried out on the siloles as organic glasses (see experimental part). The curve obtained with silole **8** shows only a first order transition at 260°C corresponding to the melting point of a crystalline solid. By contrast, the curves obtained for siloles **9** and **10** show only second order transitions (Tg) at 99°C and 96°C, respectively, which are characteristic of amorphous solids. Therefore, the introduction of either *n*-hexyl chains like in **9** or diphenylether bridges like in **10** induce the apparition of additional degrees of freedom that avoids long range molecular packing that is observed with the more rigid crystalline silole **8**.

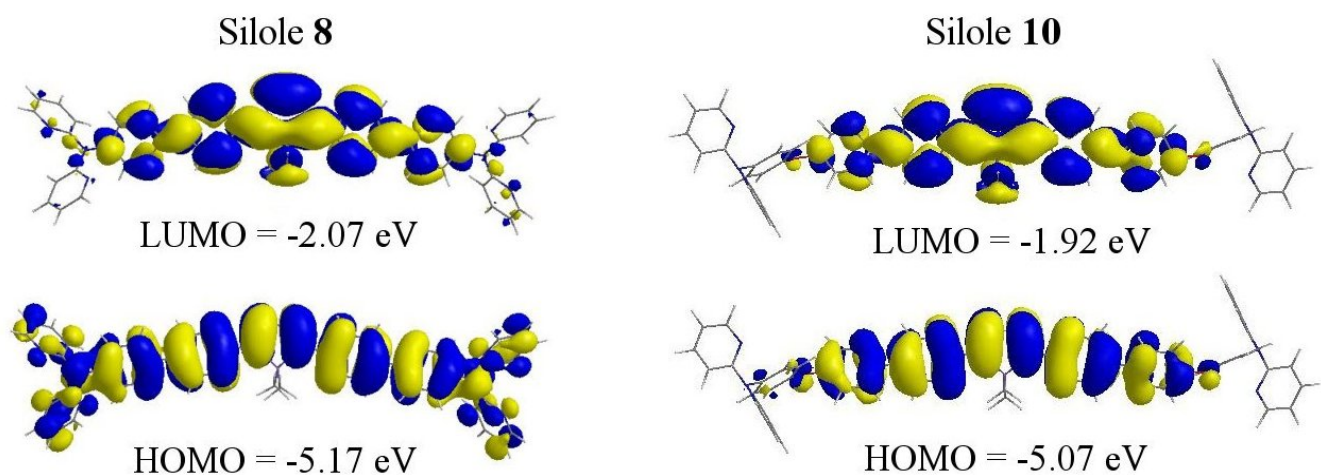

Fig. 2. B3LYP/6-31G* frontier orbitals of siloles **8** and **10**.

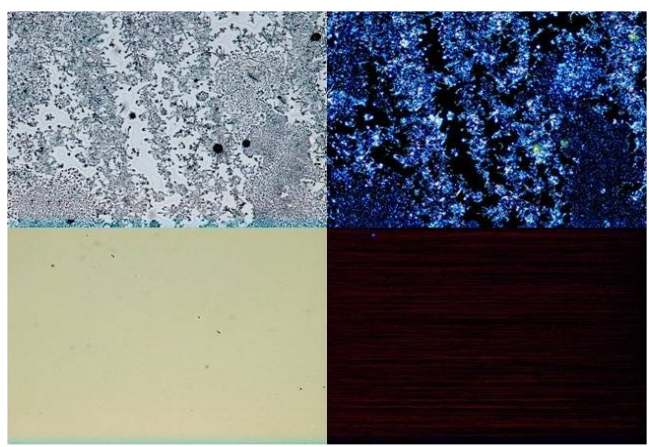

Fig. 3. On the left: Optical micrographs of vapor-deposited silole films (50 nm) morphology on Pedot-PSS/ITO substrates. On the right: Polarized optical microscopic images. These pictures are given at the top for Silole **8** and at the bottom for siloles **9** or **10.**

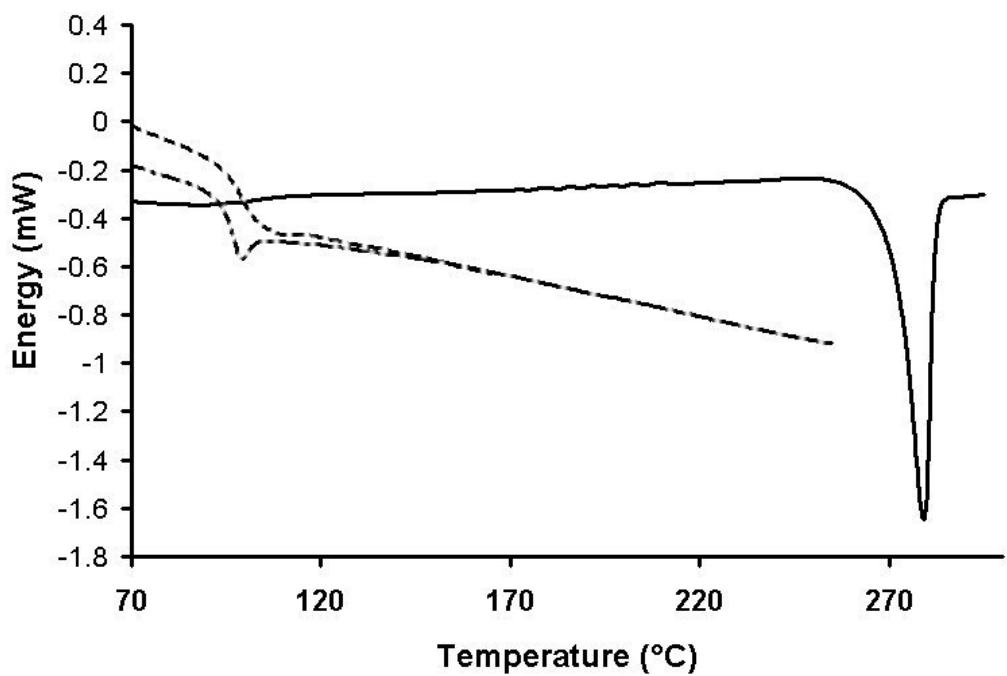

Fig. 4. DSC analysis of siloles **8** (—), **9** (---), and **10** (— ·) in the form of organic glasses.

*3.5. Electroluminescence properties*

On one hand, single layer devices were realized because it is of fundamental interest to understand the electrical behavior related to each silole. The structure of the single-layer devices has been systematically based on ITO/PEDOT for the anode and calcium for the cathode. The emissive layer, sandwiched between both electrodes, is always 50 nm thick in order to make coherent any comparisons. A, B and C are single-layer devices based on molecule **8**, molecule **9** and molecule **10**, respectively. All devices emit in the yellow-green region (Fig. 5). The emission of **10** (device D) presents a 25 nm blue shift compared to both other siloles due to the conjugation disruption and/or heteroatom effect (see above). The shape of the EL spectra do not vary with the applied voltage and lead to (X;Y) CIE 1964 chromatic coordinates such as (0.45 ; 0.53), (0.41 ; 0.56) and (0.35 ; 0.58) for A, B and C devices, respectively. Current density-voltage (*J-V*) and luminance-voltage (*L-V*) characteristics are presented in Fig. 6. Device A exhibits a high threshold voltage of 11V, twice higher than for both B and C. This voltage corresponds to the beginning of the luminance detection. Values of 90 Cd/m² at 17 V, 850 Cd/m² at 9 V and 500 Cd/m² at 9 V are reached for device A, B and C respectively. Moreover, a luminance of 100 Cd/m² is reached for device A with a current density of 200 mA/cm², whereas only 20 mA/cm² and 40 mA/cm²

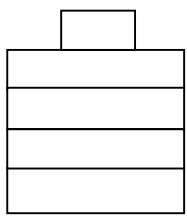

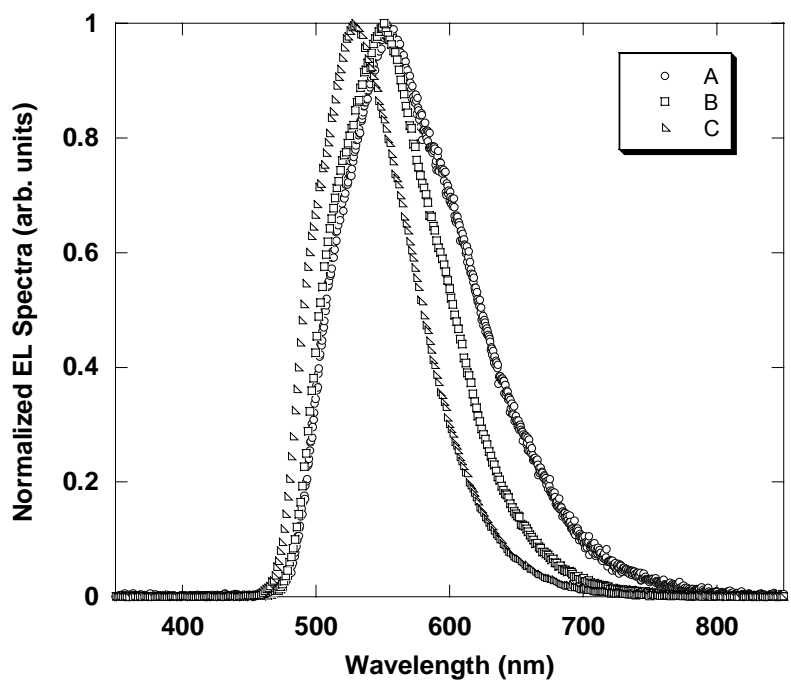

Fig. 5. Normalized electroluminescence spectra of devices A, B and C.

are necessary for devices B and C, respectively.

Device A and B are based on the electronically identical siloles **8** and **9** since the replacement of methyl groups by hexyl groups on the silicon atom didn't bring any difference, both in the shape and in the energy of the HOMO-LUMO molecular orbitals. Therefore, the difference of efficiency between these two devices may be directly connected to the difference of molecular organization in the thin film, as shown by polarized-light microscopy. Whereas their molecular and electronic structures are less similar, silole **10** (device C), posseses characteristics very close to silole **9** (device B). They have the same turn-on voltage, and the difference of luminance measured at 9 V is fairly well correlated to their quantum yield difference.[8] Actually, this result is not so surprising since injection and transport processes studies conduced on ITO/PEDOT/Silole/Ca devices have demonstrated that the hole contribution to the current was independent of the energy barrier for the electron injection.[22] In such devices, the electron current which is essencialy carried out by the silole core is much more higher than the hole current which is mainly assumed by the dipyridylamino side-groups. Theferore, the recombination efficiency of each emissive compound appears to be closely related to its molecular organization. The high current density and low luminance efficiency for device A may be thus connected to long-range molecular stacking that is characterized by the crystallization of the organic thin film. As a consequence, an excess of electron current flow is involved compared to the hole one and the recombination rate is poor. On the contrary, B and C offer better performances, indicating that either hexyl groups or diphenylether bridges avoid unfavorable π-stacking, while keeping good intermolecular orbital overlaps to support charge carrier transport.

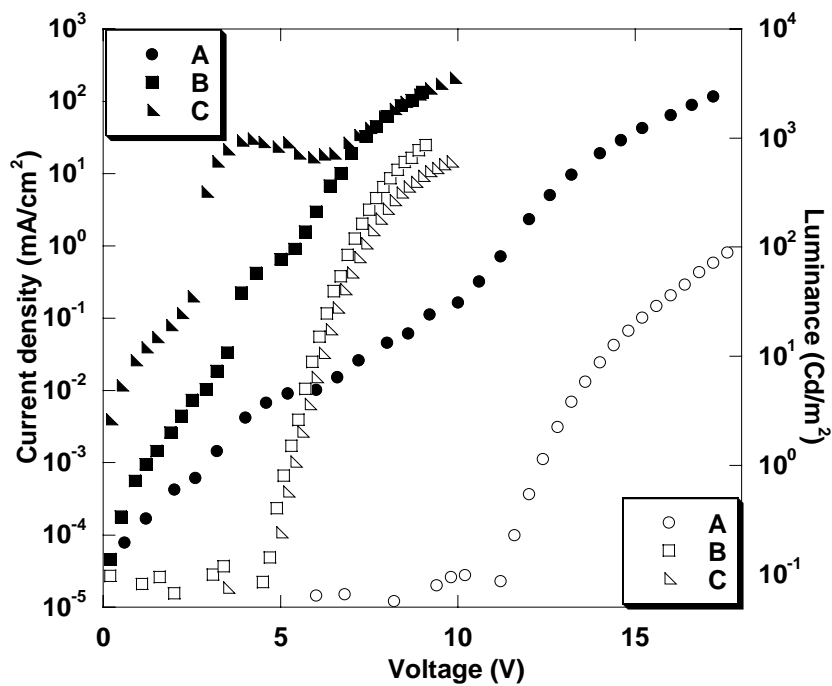

Fig. 6. Current density-voltage and luminance-voltage characteristics for single-

layer devices based on the three silole derivatives as emitting layer. Open and

**4. Conclusions**

Three new silole derivatives, especially designed for single-layer OLED structures, were presented. They were obtained by a new synthetic route and possess very strong quantum yields in solution on account to both the removal of the phenyl rings at the 3,4-positions, which are inherent to the Tamao's procedure, and to the extension of the conjugated backbone. Starting from silole **8**, which displays a strong ability to crystallize in thin films, we have chemically modified its structure to disfavor the packing of the molecule and therefore the π-stacking. For that, we have either grafted long-chain alkyls on the silicon atom to reduce the Van der Waals' interactions (silole **9**) or introduced conformational disorder by inserting an ether bridge into the conjugated rigid system (silole **10**). These modifications allowed us to obtain good quality amorphous films by vapor deposition onto PEDOT-PSS/ITO substrates. The examination of the EL properties of the devices fabricated with these compound gives evidences of the prominent role of the molecular organization on the OLED efficiency. A crystalline-like organization of the molecules allows high current density but low luminance efficiency since an excessive electron current flow is involved compared to the hole one, and the recombination rate is poor. On the contrary, disordered assemblies of molecules allows better performances by avoiding unfavourable π-stacking, while keeping good intermolecular orbital overlaps to support charge carrier transport. Thus single-layer devices based on silole **9** exhibit luminance efficiency as high as 0.75 Cd/A associated with both good chemical and film stabilities that are very promising applications for long life OLED application.

## 5. Aknowledgements

This work was supported by the French CNRS and the Université Montpellier II. PG and LA are indebted to the Région Languedoc-Roussillon for its financial support and for the award of a PhD Thesis. LH, LV and NH are grateful to the Région Aquitaine for its financial support.